\documentclass[twocolumn, 10pt, journal]{IEEEtran}
\IEEEoverridecommandlockouts
\usepackage{lineno}
\usepackage{subfigure}

\usepackage{graphicx}
\usepackage{amsmath}
\usepackage{cite}
\usepackage{float}
\usepackage{lettrine}
\usepackage{soul}
\usepackage{color}
\usepackage[table]{xcolor}
\usepackage{url}
\usepackage{diagbox}
\usepackage{multirow}
\usepackage{array}
\usepackage[ruled]{algorithm2e}
\sethlcolor{yellow}

\newcolumntype{M}[1]{>{\centering\arraybackslash}m{#1}}
\newcolumntype{N}[1]{>{\raggedright\arraybackslash}m{#1}}
\newcommand{\eat}[1]

\usepackage{tikz}
\newcommand*\circled[1]{\tikz[baseline=(char.base)]{
            \node[shape=circle,draw,inner sep=0.2pt] (char) {#1};}}

\usepackage{booktabs}

\begin{document}

\title{Quantum-Secure Microgrid
\thanks{This work was supported by the National Science Foundation under Grant ECCS-1831811.}
\thanks{Z. Tang, Z. Jiang and P. Zhang are with the Department of Electrical and Computer Engineering, Stony Brook University, Stony Brook, NY 11794, USA (e-mail: p.zhang@stonybrook.edu).}
\thanks{Y. Qin and W. O. Krawec are with the Department of Computer Science and Engineering, University of Connecticut, Storrs, CT 06269, USA.}
}

\author{\IEEEauthorblockN{Zefan Tang,~\IEEEmembership{Student Member,~IEEE,} Yanyuan Qin,~\IEEEmembership{Student Member,~IEEE,} Zimin Jiang,~\IEEEmembership{Student Member,~IEEE,} Walter O. Krawec, Peng Zhang,~\IEEEmembership{Senior Member,~IEEE}}}

\maketitle

\begin{abstract}
Existing microgrid communication relies on classical public key systems, which are vulnerable to attacks from quantum computers. This paper uses quantum key distribution (QKD) to solve these quantum-era microgrid challenges. Specifically, this paper makes the following novel contributions: 1) it offers a QKD-based microgrid communication architecture for microgrids; 2) it shows how to build a quantum-secure microgrid testbed in an RTDS environment; 3) it develops a key pool sharing (KPS) strategy to improve the cyberattack resilience of the QKD-based microgrid; and 4) it analyzes the impacts of critical QKD parameters with the testbed. Test results provide insightful resources for building a quantum-secure microgrid.
\end{abstract}


\vspace{-8pt}
\section{Introduction}
\IEEEPARstart{S}{ecuring} data transmission in microgrid is crucial for maintaining normal grid operations and achieving desirable benefits, e.g., fast recovery during a main grid blackout, improved system reliability and resilience, and economic power supply to customers~\cite{farrokhabadi2019microgrid,ren2016enabling}. Existing methods on this topic largely rely on cryptographic systems such as the Advanced Encryption Standard (AES)~\cite{van2014quantum}. AES and similar methods use a key for all encryptions within a given time period~\cite{banik2019compact}. It therefore requires that the key, which is pre-shared by two parties, has to be kept secret. This secure key distribution process is mostly achieved by public-key cryptographic methods such as the Diffie-Hellman key exchange (DH)~\cite{liskov2019enrich} and Rivest-Shamir-Adleman (RSA)~\cite{coutinho1999mathematics}.

However, the security of all classical public key systems is only guaranteed based on the assumed limits on an adversary's power. For instance, some mathematical problems such as the discrete logarithm problem~\cite{mccurley1990discrete} or the factoring problem~\cite{shor1994algorithms} cannot be effectively solved even by the fastest modern computers using any existing algorithms~\cite{yan2019logarithm}. These assumptions however are still unproven, and if proven false, the current cryptographic systems will no longer be secure~\cite{lara2019trends}.

Further, even if these assumptions remain true, the development of quantum computers will lead to security breaks~\cite{fano2019quantum,wright2019benchmarking}. Quantum computing promises to efficiently solve mathematical problems by using quantum-mechanical phenomena such as superposition~\cite{kovachy2015quantum} and entanglement~\cite{bengtsson2017geometry}. Note that although today's quantum computers are still noisy and their advent on a scale large enough to break current cryptographic systems is perhaps still decades away, their sudden appearance will leave microgrid stakeholders little time to adapt.


A potent solution to tackle this quantum-era challenge is the use of quantum key distribution (QKD)~\cite{orus2019quantum}. It uses laws of quantum mechanics to securely generate keys for two parties. Because those laws have been fairly heavily tested, they provide a more solid foundation than computational assumptions. However, although QKD has been widely applied in such areas like computer networks~\cite{hong2020quantum}, online banking~\cite{bani2019online}, and ATM transactions~\cite{cobourne2011quantum}, the microgrid community is unfortunately largely silent on the topic of developing a quantum-secure microgrid. Part of the reason for this stems from the fact that the existing QKD systems cannot be directly applied in microgrid. With multiple communication channels and different transmission requirements existing in microgrid, it was unclear how QKD performs and whether it is applicable under various circumstances. A real-time QKD-integrated microgrid simulation testbed for evaluating the performance of the QKD-based microgrid is essential but does not yet exist.

Further, the key generation speed in a QKD system is affected by a number of variables like the distance between two communicating parties and the noise, which can be either natural or caused by an adversary, on quantum optic equipment. A large distance or a strong attack on the QKD equipment can reduce this speed, detrimentally causing keys to be exhausted. A proper strategy is significantly needed to enhance the cyberattack resilience for the system.

To bridge the gaps, in this paper, we develop a QKD-integrated microgrid testbed in Real Time Digital Simulator (RTDS). Key components like hardware connection, communication network, and QKD integration are designed and presented in detail. This is an important step towards constructing a realistic QKD-enabled microgrid in practice. The real-time communication between the RTDS simulator and a remote server enabled by the QKD algorithm is the salilent feature of this testbed. Main contributions of this paper are fourfold:
\begin{itemize}
    \item A novel QKD-enabled communication architecture is devised for microgrids.
    \item A QKD-integrated microgrid testbed is built in RTDS. Key components like hardware connection, communication network, and QKD integration are presented.
    \item A key pool sharing (KPS) strategy is designed to further enhance the system's resilience to cyberattacks.
    \item The impacts of critical QKD parameters like quantum fiber length, data transmission speed, attack level, and detection efficiency are evaluated with the testbed.
\end{itemize}

The rest of this paper is organized as follows: Section II describes quantum communication, and presents the design of the QKD-based microgrid architecture and KPS strategy. Section III elaborates the testbed design. Our evaluation results are reported in Section IV, and Section V concludes the paper.

\section{Quantum-Secure Microgrid Enabled by QKD}
In this section, we first briefly introduce the topic of quantum communication, including the quantum states, the general setting of a QKD system, and a practical QKD protocol. We then present the benefits of using QKD for microgrids and propose a QKD-based microgrid communication architecture. At the end, we present our novel KPS strategy for improving the system's cyberattack resilience.

\vspace{-4pt}
\subsection{Quantum Communication}
Unlike classical secure key distribution systems that rely on mathematical assumptions, quantum communication utilizes a radically different foundation: the uncertainty principle of quantum physics. In this subsection, a brief overview of quantum communication is presented, including an introduction to quantum states, the general setting of a QKD system, and the practical QKD protocol used in this paper.

\subsubsection{Quantum States}
Instead of using binary bits to encode information as in classical communication systems, quantum communication utilizes quantum states, or ``qubits''. A qubit is a two-state quantum-mechanical system, whose state is commonly represented by the spin of an electron or the polarization of a photon. Unlike a binary bit, which has to be in one state or the other, a qubit can be in a coherent superposition of both states~\cite{trauzettel2007spin}. For QKD systems, photons are the primary practical implementation of qubits. For the QKD system we consider, the polarization of the photon will be used to encode a quantum state. We will consider two \emph{Bases}, namely horizontal polarization (denoted the $Z$ basis later) and diagonal polarization (denoted the $X$ basis later). If a source and its receiver both operate in the same basis, information can be transmitted deterministically; however, if different bases are used, the information received will be uncorrelated with the transmitted information. The security of a QKD protocol, in a way, takes advantage of this: by encoding a classical bit string using different, randomly-chosen bases, an adversary who is unaware of the basis choice can never be truly certain of the information being transmitted. Furthermore, any attempt to actually learn this information causes noise in the quantum channel which may be detected by the users later.

\subsubsection{General Setting}
The general setting of a QKD-based communication system consists of a quantum channel and a classical one. The quantum channel allows two parties to share quantum signals for creating a secure and secret key. With the created key, the information to be transmitted is encrypted and later decrypted over the classical channel. The \emph{key generation rate} of a QKD protocol is an important statistic and is affected by numerous parameters, most importantly the noise in the quantum channel (caused, perhaps, by an adversary or natural noise) and the distance between the two parties.

An important and unique property of QKD is that the two parties can detect when an eavesdropper is trying to gain knowledge of the key. This is due to the quantum-mechanical property that measuring an unknown quantum state will, in general, change that state. This ensures that a non-secret key will never be used, making QKD-based encryption and authentication theoretically secure. It is worth noting that QKD is only used to generate the key in the quantum channel; the message data is still transmitted using classical encryption methods over the classical channel. In reality, QKD can be associated with either one-time pad (OTP) or symmetric key algorithms such as AES.

\subsubsection{Practical QKD Protocol}
Different protocols have been proposed to implement QKD such as the well-known BB84, decoy-state, six-state, Ekert91, and BBM92. In this paper, we consider a practical decoy-state QKD protocol~\cite{pirandola2019,lim2014concise}. This protocol has been one of the most widely used schemes in the QKD community because of its ability to tolerate high channel loss and to operate robustly even with today's hardware. Its security and feasibility have been well-demonstrated by several experimental groups, and theoretical security analyses including the evaluation of concise and tight finite-key security bounds have also been provided.

The idea of this protocol is as follows: The information is encoded into qubits and then sent out by one party, commonly named Alice, using weak coherent laser pulses. With today's technology, the production of a single qubit is not practical; instead, weak coherent laser pulses are used. However, these pulses contain, with non-zero probability, multiple qubit signals that would cause a break in security. To tackle this challenge, the decoy-state protocol varies the intensity of each laser pulse  randomly using one of three intensities $k_1$, $k_2$ and $k_3$, which are the intensities of the signal state, decoy state and vacuum state, respectively. Two bases $X$ and $Z$ are selected with probabilities $p_x$ and $1-p_x$, respectively. Recall that these bases refer to the polarization setting of the qubit. The other party, named Bob, measures the qubits by randomly selecting bases from $X$ and $Z$. If Alice and Bob choose the same basis, they share information since sending and receiving qubits in the same basis, as mentioned, leads to a deterministic outcome; otherwise, the iteration is discarded. By repeating this numerous times, the two parties share a so-called \emph{raw-key}, which is partially correlated and partially secret. Error correction is then performed (leaking additional information to the adversary which must be taken into account) followed by privacy amplification, yielding a secret key of size $\ell$. One is often interested in the key generation rate $\ell/N$, where $N$ is the number of signals needed to produce a raw-key of sufficient size for generating the secret key of size $\ell$.

Specifically, the procedures of this protocol are described below~\cite{lim2014concise}:

\begin{itemize}
    \item Step 1: Preparation. Alice selects a bit value from 0 and 1 uniformly at random; a basis from $X$ and $Z$ with probabilities $p_x$ and $1-p_x$, respectively; and an intensity from $k_1$, $k_2$ and $k_3$ with probabilities $p_{k_1}$, $p_{k_2}$ and $p_{k_3}=1-p_{k_1}-p_{k_2}$, respectively. Based on the selected values, Alice prepares a laser pulse and sends it to Bob through the quantum channel. Note that Alice sends Bob the information qubit by qubit.
    \item Step 2: Measurement. When Bob receives the qubits from Alice, for each qubit he randomly selects a basis from $X$ and $Z$ with probabilities $p_x$ and $1-p_x$, respectively. He then decodes the qubit using the selected basis.
    \item Step 3: Basis reconciliation. Alice announces the basis and intensity choices, and Bob announces the basis choices.  Note that, this is done \emph{after} the qubits are received by Bob. Due to the no-cloning theorem~\cite{nagata2019no}, this information is no longer helpful to the eavesdropper as she could not copy the originally-sent qubits to measure now. The raw-key bits are extracted from the events where Alice and Bob both select the $X$ basis.
    \item Step 4: Generation of the raw key and the error estimation. Alice and Bob generate a raw key pair ($X_A, X_B$) by using all events where they chose the $X$ basis. Events from the $Z$ basis are used for quantum error estimation.
    \item Step 5: Post-processing. Alice and Bob execute an error correction algorithm trying to correct for a predetermined error rate. To ensure that the error correction has been successful and that they have shared identical keys, they perform an error verification using hash functions. Finally, they perform a privacy amplification to extract a secret key pair.
\end{itemize}

Once all the post-processing procedures have been successfully completed, the key is established and can be used by Alice and Bob. The length $\ell$ of the extracted secret key can be obtained in the following way \cite{lim2014concise}:

\vspace{-12pt}
\begin{small}
\begin{equation}
   \ell=\lfloor \xi_{X,0}+\xi_{X,1}-\xi_{X,1}h(\phi_X)-\lambda_{ec}-6\log_2 \frac{21}{\varepsilon_{s}}-\log_2 \frac{2}{\varepsilon_{c}} \rfloor,
    \label{equ:length}
\end{equation}
\end{small}where $h(x)=-x\log_2x-(1-x)\log_2(1-x)$ is the binary entropy function. $\xi_{X,0}$, $\xi_{X,1}$, and $\phi_X$ are the number of vacuum events, the number of single-photon events, and the phase error rate associated with the single-photon events in $X_A$, respectively. $\varepsilon_{c}$ is the probability that the keys extracted by the two parties are not identical, and $\varepsilon_{s}$ is the user-specified maximum failure probability. $\lambda_{ec}$ specifies how much information leaked during error correction. It is set to $n_X\eta_{ec}h(\phi_{X})$, where $n_X$ is the size of the raw key $X_A$, and $\eta_{ec}$ is the error-correction efficiency.

The above parameters cannot be directly observed; however, by using the decoy-state protocol, they can be bounded.  Let $n_{X,k}$ be the number of $X$ signals received using intensity $k$.  Then, of course, $n_X$, the size of the raw key, is simply the sum of all $n_{X,k}$ over all the intensities used. Basically, the number of vacuum events in $X_A$, $\xi_{X,0}$, satisfies
\begin{equation}
    \xi_{X,0}\geq \chi_0 \frac{k_2 n_{X,k_3}^{-}-k_3 n_{X,k_2}^{+}}{k_2-k_3},
    \label{sx0}
\end{equation}
where $\chi_n$ is the probability that Alice sends a $n$-photon state.  This value, using a weak-coherent laser, follows a Poisson distribution and is found to be:
\begin{equation}
    \chi_n=\sum_{k\in \{k_1, k_2, k_3\}} e^{-k}k^np_k/n!,
    \label{tau}
\end{equation}
and
\begin{equation}
    n_{X,k}^{\pm} = \frac{e^k}{p_k} (n_{X,k} \pm \sqrt{\frac{n_X}{2} \ln \frac{21}{\varepsilon_{s}}} ), \ \forall k \in \{k_1, k_2, k_3\}.
    \label{nxk}
\end{equation}

The number of single-photon events in $X_A$, $\xi_{X,1}$, satisfies
\begin{equation}
    \xi_{X,1}\geq \frac{\chi_1k_1[n_{X,k_2}^{-}-n_{X,k_3}^{+}-\frac{k_2^2-k_3^2}{k_1^2}(n_{X,k_1}^{+}-\frac{\xi_{X,0}}{\chi_0})]}{k_1(k_2-k_3)-k_2^2+k_3^2}.
    \label{sx1}
\end{equation}

Similarly, by using (\ref{sx0})-(\ref{sx1}) with statistics from the basis $Z$, the number of vacuum events in $Z_A$, $\xi_{Z,0}$, and the number of single-photon events in $Z_A$, $\xi_{Z,1}$, can also be obtained.

The phase error rate of the single-photon events in $X_A$, $\phi_X$, satisfies \cite{fung2010practical}, 
\vspace{-8pt}
\begin{equation}
    \phi_X \leq \frac{\delta_{Z,1}}{\xi_{Z,1}}+f(\varepsilon_{s}, \frac{\delta_{Z,1}}{\xi_{Z,1}}, \xi_{Z,1}, \xi_{X,1}),
    \label{phix}
    \vspace{-8pt}
\end{equation}
where
\vspace{-4pt}
\begin{equation}
    f(a,b,c,d)=\sqrt{\frac{(c+d)(1-b)b}{cd \log 2}\log_2(\frac{c+d}{cd(1-b)b} \frac{441}{a^2})},
    \label{gamma}
    \vspace{-4pt}
\end{equation}
and $\delta_{Z,1}$ is the number of bit errors of the single-photon events in $Z_A$. It is given by
\vspace{-8pt}
\begin{equation}
    \delta_{Z,1} \leq \chi_1 \frac{m_{Z,k_2}^+ - m_{Z,k_3}^-}{k_2-k_3},
    \vspace{-8pt}
\end{equation}
where
\begin{equation}
    m_{Z,k}^{\pm} = \frac{e^k}{p_k}(m_{Z,k}\pm \sqrt{\frac{m_Z}{2}\ln \frac{21}{\varepsilon_{s}}}), \ \forall k \in \{ k_1, k_2, k_3 \},
\end{equation}
and $m_Z=\sum_{k\in \{ k_1, k_2, k_3 \}}m_{Z,k}$. Here, $m_{Z,k}$ is the number of error events in the $Z$ basis.  For more details on how the size of the secret key $\ell$ is computed through the above equations, readers are referred to \cite{lim2014concise}.

The above equations are general for any observations.  For our simulation, we will assume a standard fiber channel and practical settings for devices.  In this case, the probability of having a bit error for intensity $k$, $b_{k}$, is as follows~\cite{eraerds2010quantum}:
\begin{equation}
    b_{k}=p_{dc}+e_{mis}(1-e^{-\eta_{tr}k})+\frac{p_{ap}r_k}{2},
    \label{probiterror}
\end{equation}
where $p_{dc}$ and $p_{ap}$ are the dark count probability and the after-pulse probability, respectively. $e_{mis}$ is the error rate due to optical errors. $\eta_{tr}$ is the transmittance that is related to the fiber length $L$ as follows:
\begin{equation}
    \eta_{tr}=10^{-0.2L/10},
    \label{etach}
\end{equation}
where the fibers are assumed to have an attenuation coefficient of 0.2 dB/km. In (\ref{probiterror}), $r_k$ is the expected detection rate (excluding after-pulse contributions), and can be calculated as follows:
\begin{equation}
    r_k=1-(1-2p_{dc})e^{-\eta_{ch}\eta_{Bob}k},
    \label{Dk}
\end{equation}
where $\eta_{Bob}$ is Bob's detection efficiency.

In this paper, the initial values of the parameters from (\ref{equ:length})-(\ref{Dk}) are given in Table~\ref{tab:paraQKD}.

\begin{table}
  \caption{Initial values of the parameters in the QKD system}
  \vspace{-4pt}
  \centering
\begin{tabular}{c c c c c c}
  \hline
   $k_1$ & $k_2$ & $k_3$ & $p_{k_1}$ & $p_{k_2}$ & $p_{k_3}$ \\
   0.4 & 0.1 & 0.007 & 1/3 & 1/3 & 1/3  \\
   \hline
    $n_X$ & $p_x$ & $p_{dc}$ & $p_{ap}$ & $\eta_{Bob}$ & $e_{mis}$ \\
   10$^7$ & 0.8 & 6$\times$10$^{-7}$ & 4$\times$10$^{-2}$ & 0.1 & 5$\times$10$^{-4}$ \\
   \hline
   $L$ (km) & $\eta_{ec}$ & $\varepsilon_{c}$ & $\varepsilon_{s}$ & & \\
   5 & 1.16 & 10$^{-11}$ & 10$^{-11}$ & & \\
   \hline
\end{tabular}
\label{tab:paraQKD}
\vspace{-4pt}
\end{table}

\subsection{Benefits of Using QKD for Microgrids}

QKD has been envisioned as one of the most secure and practical instances of quantum cryptography. Specifically, using QKD provides the following benefits for microgrid:

\begin{itemize}
    \item The key generated by QKD in microgrid is almost impossible to steal even in the face of an adversary with infinite supplies of time and processing power.
    \item QKD is a particularly good method for producing a long random key, which makes the OTP much more realistic in practice. When QKD is combined with one-time pads (OTPs), both the key generation and the encryption processes are unconditionally secure.
    \item A QKD-enabled microgrid is able to detect the presence of an eavesdropper trying to gain knowledge of the key, whereas existing communication systems without this ability will inevitably require extra detection mechanisms.
    \item QKD systems have the advantage of being automatic compared with manually distributing keys in microgrid.
\end{itemize}

\vspace{-10pt}
\subsection{Quantum-Secure Microgrid Communication Architecture}

\begin{figure}[t]
\centering
\includegraphics[width=0.48\textwidth]{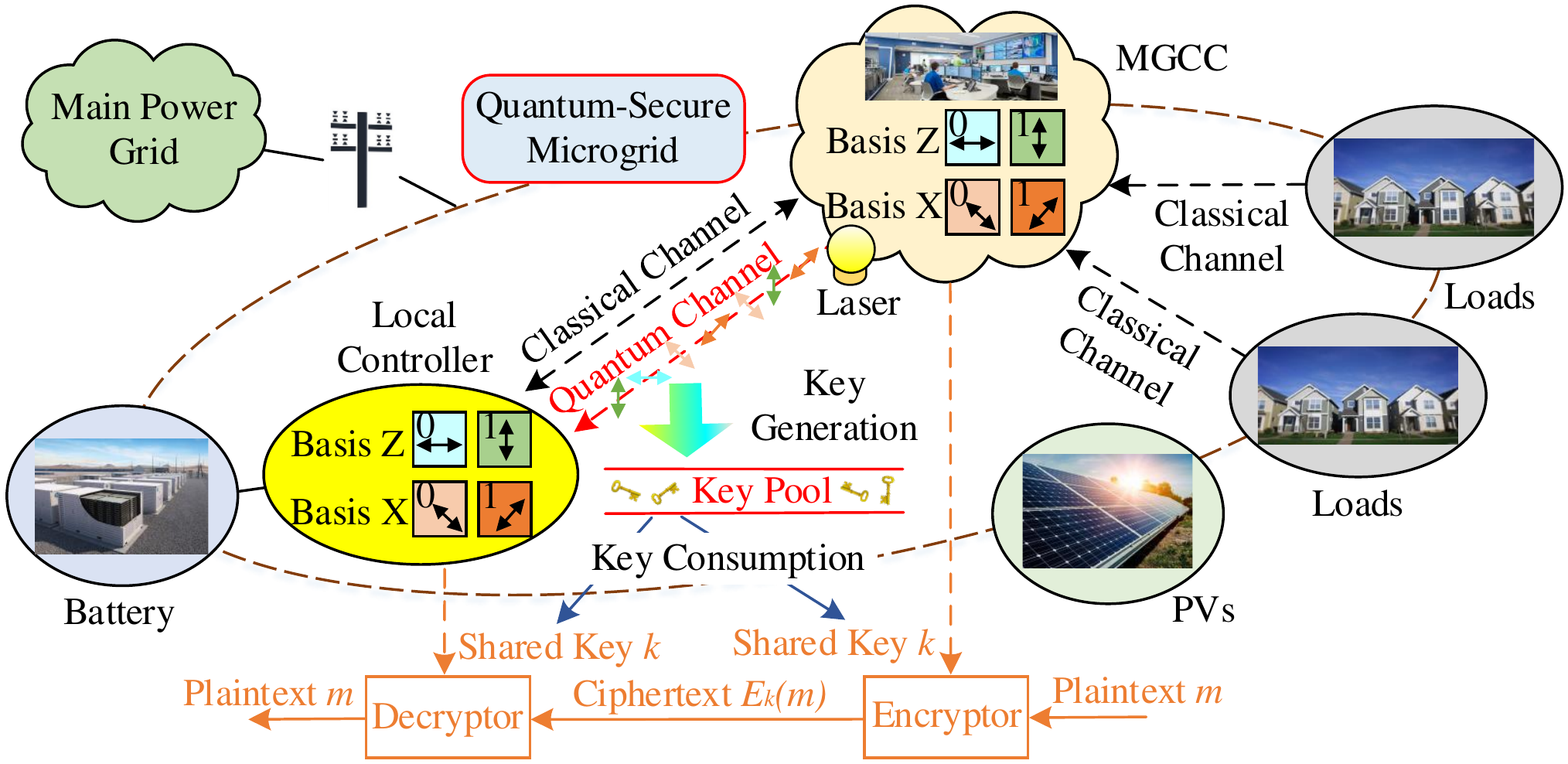}
\caption{\label{fig:QKDMG}QKD-enabled quantum-secure microgrid communication architecture.}
\vspace{-12pt}
\end{figure}

Given the great benefits described above, we present a QKD-based communication architecture for microgrids. As illustrated in Fig.~\ref{fig:QKDMG}, the microgrid control center (MGCC) collects data from different loads and sends control signals to local controllers. As building a quantum channel is really costly, it is practical and reasonable to implement QKD for only those critical communication channels in microgrid. In this study, without loss of generality, a QKD-based quantum channel is built between the MGCC and the local controller for a battery's storage. This battery uses a P-Q control to adjust its power output based on real and reactive power references received from the MGCC. It is worth noting that, QKD is only used for generating keys for two parties in an unconditional secure way; the data encryption process is still achieved using classical cryptographic methods such as AES or OTP. Using AES to encrypt data is considered quantum-secure, as long as the key used for this process is secure~\cite{chen2016report}. OTP is even more secure (or more accurately, unconditionally secure), because it uses a random key only once and then discards the key. But this requires that the key be as long as the plaintext. Keys generated by a QKD link are stored in a key pool, and when there is a need to transfer data, a certain number of key bits are extracted for encryption and decryption purposes. 

To properly integrate QKD into microgrid, a critical concern is key generation speed in a QKD system. It has to be larger than the frequency of data transmission to guarantee there are \textit{always} enough keys in the key pool. Critical QKD parameters that affect the key generation speed include quantum fiber length, attack level, and receiver's detection efficiency.

Different with other applications where there is no strict requirement on the frequency of data transmission, microgrid often needs a high frequency of continuous data transmission to accommodate fast and dynamic changes typically caused by customers or various distributed energy resources (DERs). Thus, before constructing a real QKD system in microgrid, building a real-time simulation testbed to evaluate the performance of the QKD-enabled microgrid under different circumstances is an important step. In this paper, we show in detail how to build a QKD-integrated microgrid testbed in RTDS, a real-time power system simulator. To maintain normal operations of the QKD-enabled microgrid when the key bits in a key pool are used up (this may be caused by increased data transmission frequency or a strong attack), we further develop a key pool sharing (KPS) strategy.

\vspace{-10pt}
\subsection{The KPS Strategy}

The idea of this strategy is as follows: The MGCC establishes multiple quantum channels with local controllers and uses separate key pools to store keys. Key pools can share keys with each other, meaning that, when the number of key bits in one key pool is below a pre-determined threshold, a certain number of key bits can be shared from other key pools.  

\begin{figure}[t]
\centering
\includegraphics[width=0.48\textwidth]{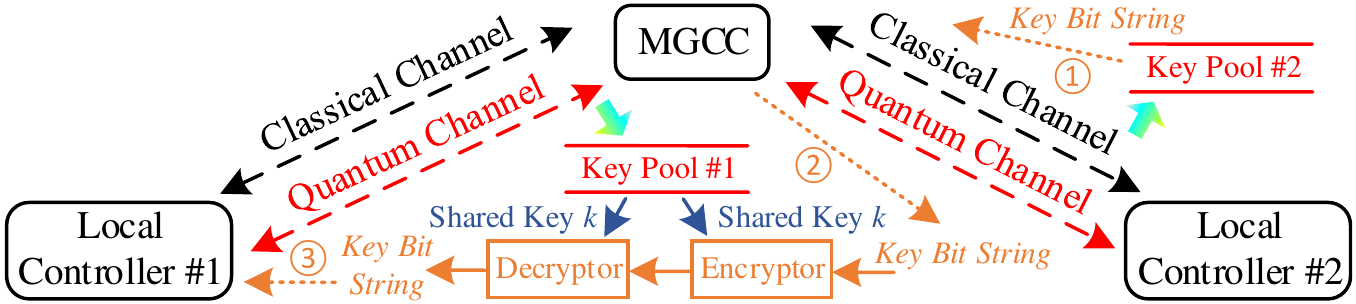}
\caption{\label{fig:KPS}An example of the KPS strategy.}
\vspace{-8pt}
\end{figure}

An example of the KPS strategy is illustrated in Fig.~\ref{fig:KPS}, where two quantum channels are established between the MGCC and two local controllers. When the number of key bits in key pool \#1 is lower than a threshold, for instance, a string of key bits is extracted from key pool \#2 by the MGCC (represented in \circled{1} in Fig.~\ref{fig:KPS}). This key bit string is then used as plaintext (represented in \circled{2} in Fig.~\ref{fig:KPS}), encrypted by the MGCC via a key extracted from key pool \#1 (note that there are still some key bits left in key pool \#1), and sent to local controller \#1. Local controller \#1 uses the same key from key pool \#1 to decrypt the received message and obtains the key bit string (represented in \circled{3} in Fig.~\ref{fig:KPS}). In this way, a string of key bits is transferred from key pool \#2 and is securely shared between the MGCC and local controller \#1. Although this distribution of keys through AES loses information-theoretic security, it is still better than relying on public key systems, because, as mentioned, AES is considered quantum-secure as long as the key used for the encryption is secure~\cite{chen2016report}.  Note that, unlike an alternative approach employing AES keys for actual data transmission (changing the key every $n$ seconds), our KPS system has the advantage that information theoretic OTP may be used up until the last $128$ or $256$ bits are available maximizing security of the overal system (switching to computational security only as a last-resort).

\textit{\textbf{Overhead analysis:}} The communication and computation overheads of our KPS strategy are negligible. Assuming the microgrid control signals with a total size of 200k bits that need to be transmitted within 20 seconds, then 200k bits of quantum keys are used to encrypt the data. The required bandwidth for transmitting those key bits from the MGCC to a local controller is therefore only 10 Kbps, which is far less than the link capacity of a common switch (i.e., 1 Gbps). On the other hand, practical encryption schemes such as 128-bit AES can be utilized to transmit quantum keys, where only a few key bits are consumed for encrypting a large number of bits (e.g., 128 bits for a 1500-byte packet). The processing time of the 128-bit AES encryption with the current computing hardware is small. A commercial server with four cores could process AES data with a speed up to 2,804 MB/s~\cite{aes}.

\section{Quantum-Secure Microgrid Test Environment}


\subsection{High-Level Design}

\begin{figure}[t]
\centering
\includegraphics[width=0.48\textwidth]{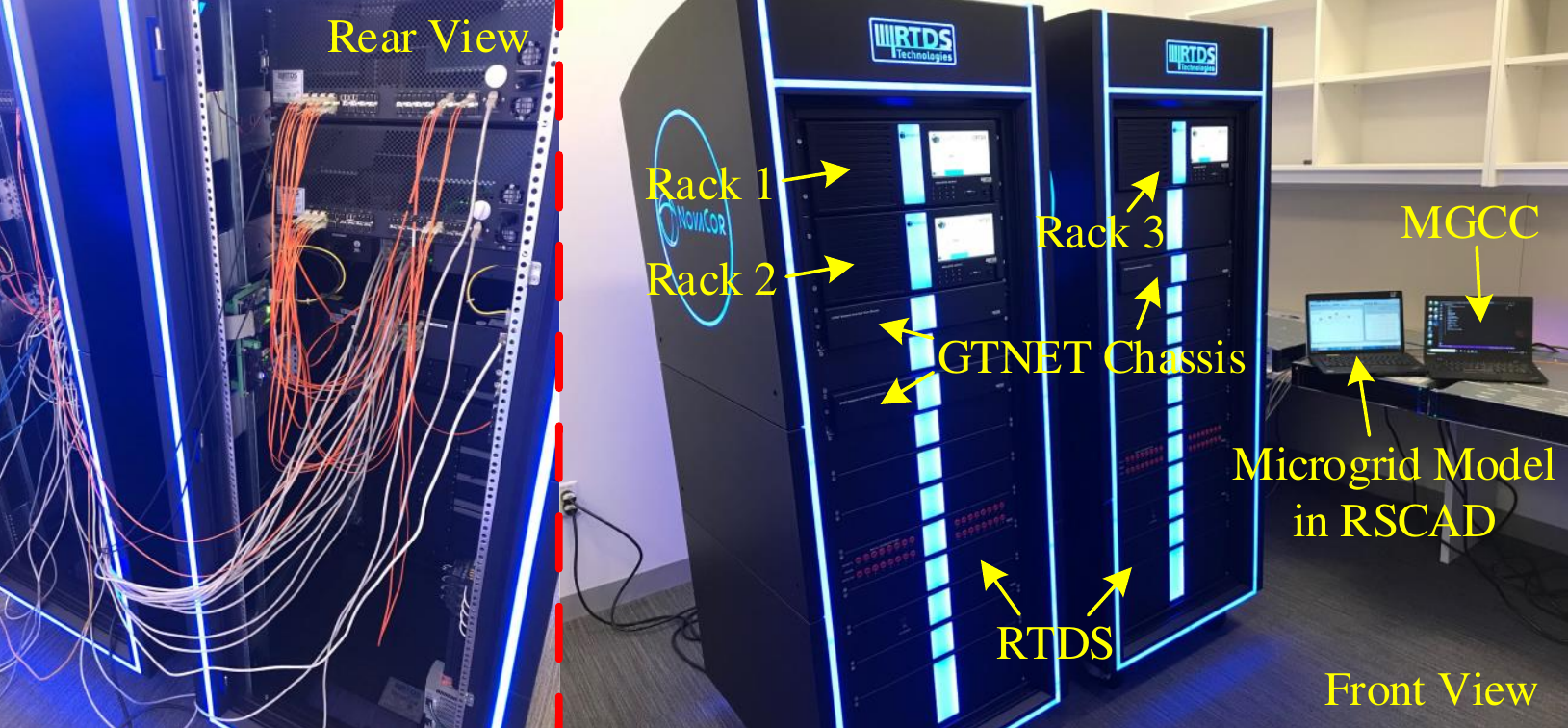}
\caption{\label{fig:testbed}Testbed setup for a quantum-secure microgrid in RTDS environment.}
\vspace{-12pt}
\end{figure}

The test environment is illustrated in Fig.~\ref{fig:testbed}. Specifically, the microgrid model is developed and compiled in RSCAD, a power system simulation software designed to interact with the RTDS simulation hardware. The RTDS in our testbed consists of three racks, which can be either used separately for small-scale power systems or combined together to provide more cores for a large-scale system. In our simulation, rack 2 is utilized to simulate the microgrid model in real-time, where the four cores in that rack (running at 3.5 GHz) are sufficient to provide high fidelity for test results in this paper.

The measurements from the RTDS simulator are transmitted through a GTNETx2 card and sent to the MGCC via a communication network.
The GTNETx2 card can either receive data from the RTDS and send it to external equipment, or it can receive data from the network and send it back to the RTDS, depending on whether the GTNETx2 card was designed to be in sending or receiving mode. The MGCC runs on a remote server, which can receive load measurements from and send signals back to RTDS with a 1 Gbps Ethernet connection.

\begin{figure}[t]
\centering
\includegraphics[width=0.48\textwidth]{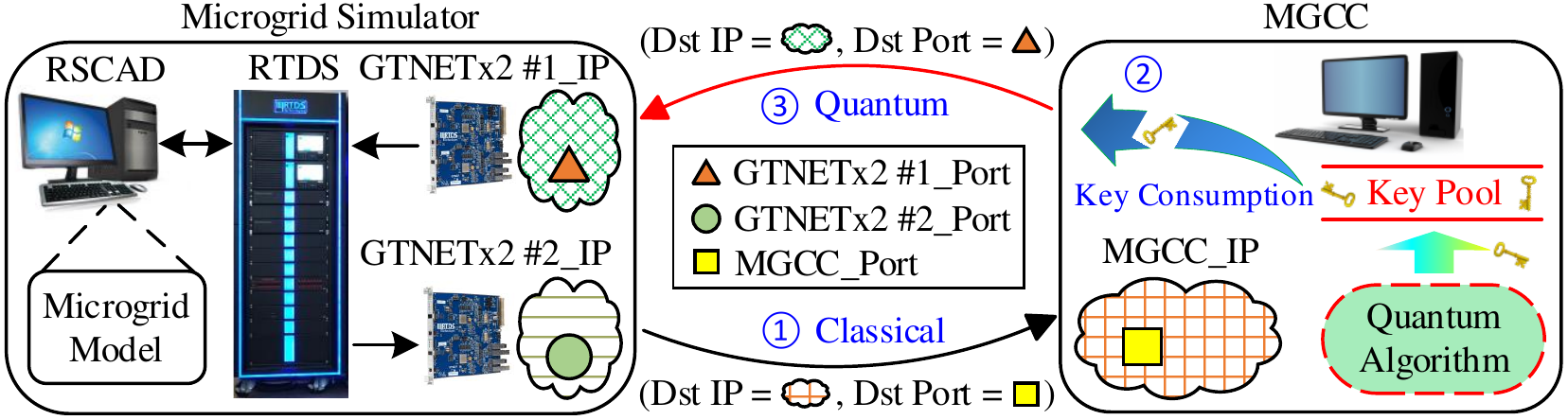}
\caption{\label{fig:Network}High-level design of the quantum-secure microgrid testbed.}
\vspace{-8pt}
\end{figure}

The high-level design of the testbed is illustrated in Fig.~\ref{fig:Network}. Two GENETx2 cards are utilized for the purpose of network communication. It should be noted that, although only one quantum channel is established in this case, the principle can be easily extended to cases with multiple quantum channels. GTNETx2 card \#2 is used to transmit data from the RTDS to the MGCC, which models the classical communication (represented in \circled{1} in Fig.~\ref{fig:Network}) in real-time, i.e., collecting load measurements to MGCC as shown in Fig.~\ref{fig:QKDMG}. When the data is received by the MGCC, an analysis of the data is conducted, and proper control signals are sent to the local controller. Before a control signal is sent out, a key with the same length is extracted from the key pool. This process (represented in \circled{2} in Fig.~\ref{fig:Network}) succeeds only when there are enough key bits in the key pool.

GTNETx2 card \#1 is utilized to receive signals from the MGCC (represented in \circled{3} in Fig.~\ref{fig:Network}) and transfer them to the RTDS. The simulation results with the updated control signals are demonstrated in RSCAD. Note that the QKD system is modeled through an algorithm formulated from (\ref{equ:length})-(\ref{Dk}). Keys are continuously generated by the QKD algorithm, and are stored in a key pool. This real-time communication between the RTDS microgrid simulator and the MGCC using the QKD algorithm is the salient feature of this testbed.

\vspace{-10pt}
\subsection{QKD-Based Microgrid Communication Network}

\begin{figure}[t]
\centering
\includegraphics[width=0.48\textwidth]{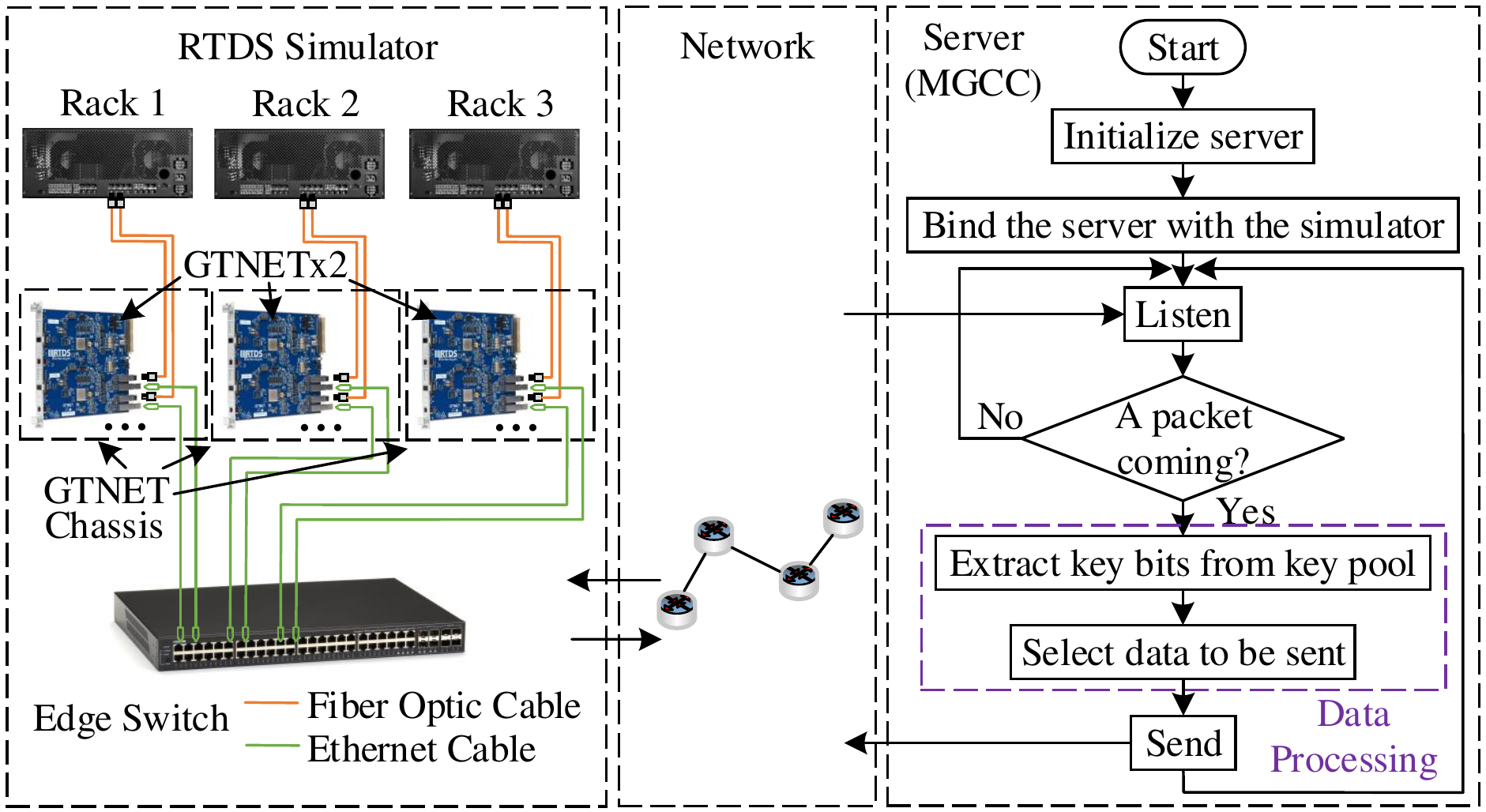}
\caption{\label{fig:FlowChart}The network connection of key components in the RTDS simulator and a flow chart of the algorithm running in the MGCC.}
\vspace{-12pt}
\end{figure}

The network connection of key components in the RTDS simulator and a flow chart of the algorithm running in the MGCC are illustrated in Fig.~\ref{fig:FlowChart}. As shown on the left side of Fig.~\ref{fig:FlowChart}, each RTDS rack is connected to one or more GTNETx2 cards using fiber optic cables. All the GTNETx2 cards are connected with an edge switch through Ethernet cables to transmit and receive data over the network. The User Datagram Protocol (UDP) is used in our simulation.

From the MGCC side, as shown on the right side of Fig.~\ref{fig:FlowChart}, the server enters the $listening$ mode after being connected to the simulator. At this stage, the server is receiving any UDP packet whose destination IP and port match those of the server, respectively. Once a packet arrives, a quantum key with the same length of the received data, i.e., 64 bits in this paper, is extracted from the key pool, and corresponding control signals are generated. The server then enters the $sending$ mode and starts to send out control signals whose destination IP and port are the IP and port of GTNETx2 card \#1 in the RTDS simulator (see Fig.~\ref{fig:Network}), respectively. After controller signals are sent out, the server goes back to the $listening$ mode.

\vspace{-8pt}
\subsection{Microgrid Modeling and Simulation}

\begin{figure}[t]
\centering
\includegraphics[width=0.44\textwidth]{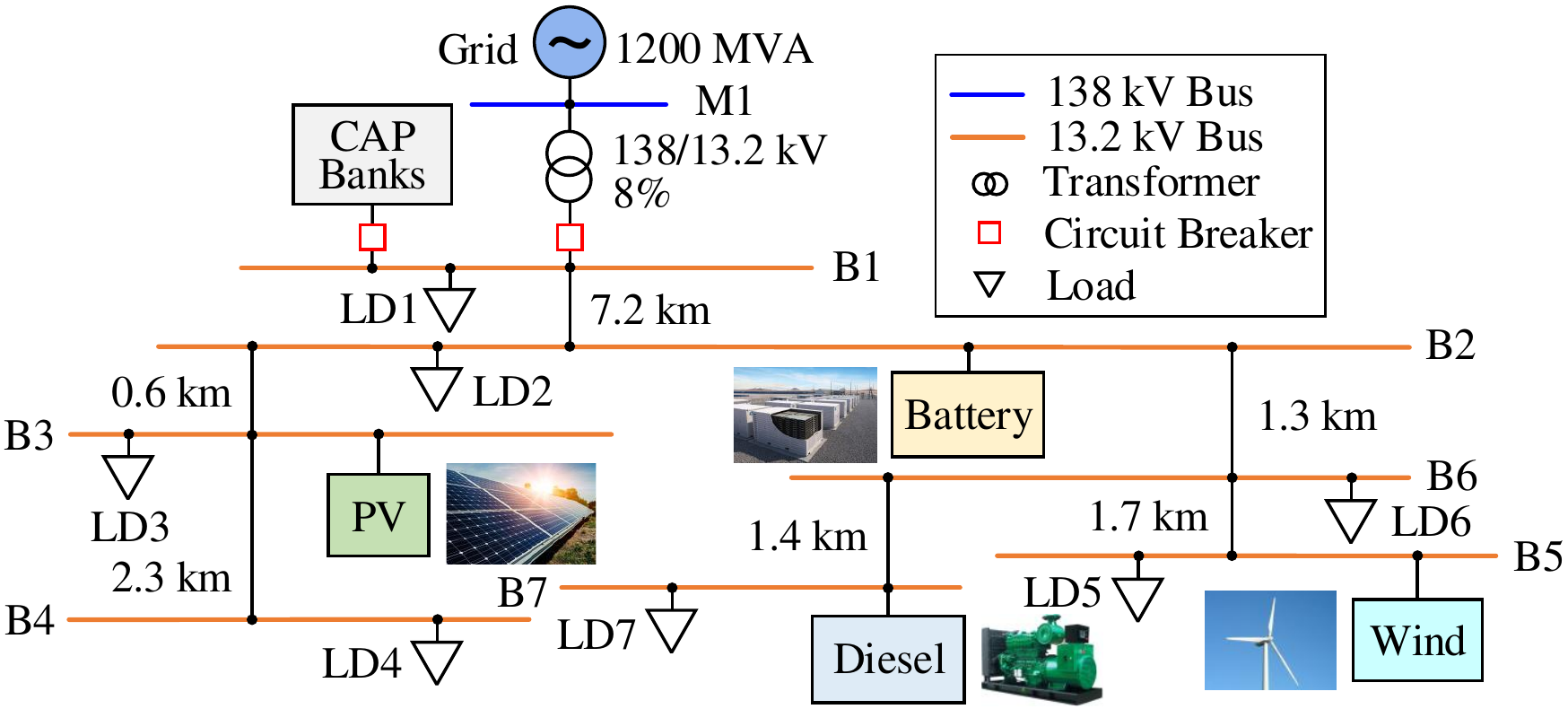}
\caption{\label{fig:MGModel}One-line diagram of the microgrid model.}
\vspace{-12pt}
\end{figure}

A typical microgrid system shown in Fig.~\ref{fig:MGModel} is used to evaluate the performance of the QKD-enabled quantum-secure microgrid in this study. This system is based on a medium-voltage microgrid from \cite{onyinyechi2015real} with a battery and communication channels added. The buses within the microgrid are rated at 13.2 kV, and the microgrid is connected to the 138 kV main grid through a 138/13.2 kV transformer and a circuit breaker. The microgrid can operate either in islanded mode or in grid-connected mode depending on the state of the circuit breaker. The transformer is $\Delta-Y$ connected and rated at 25 MVA with a 8\% impedance.

The DERs in the microgrid include a 5.5 MVA diesel generator, a 1.74 MW PV system, and a 2 MW doubly-fed induction generator wind turbine system. The diesel generator uses the droop control to regulate the microgrid frequency in islanded operation and to provide real and reactive powers in both grid-connected and islanded modes. The PV system and wind turbine both use the MPPT control to maximize their power outputs. Three switched capacitors are connected at bus 1 to facilitate voltage synchronization in the microgrid.

A lithium-ion battery storage is further connected at bus 2 to provide a backup power supply and store extra energy when the microgrid is in islanded operation. The battery model consists of 250 stacks connected in parallel with each one having 250 cells in series. A single cell has a capacity of 0.85 AH, and the initial state of charge in a single cell is set at 85\%. A P-Q control is designed to regulate the output power of the battery, the value of which is determined by the real and reactive power references transferred from the MGCC via a communication channel. The initial values of the real and reactive power references are both set at zero.

The resistance and inductance of a unit length of the lines in the microgrid are 0.2322 $\Omega$/km and 2.355$\times 10^{-3}$ H/km, respectively, and the lengths of the lines are given in Fig.~\ref{fig:MGModel}. For more details on the microgrid, readers are referred to \cite{onyinyechi2015real}.

\section{Experimental Results}
In this section, we evaluate the performance of the QKD-based microgrid communication with our hardware testbed. The results include 1) a comparison of the performance with different data transmission speeds, 2) the impact of cyberattacks on the microgrid, 3) an evaluation of the key generation speed under different fiber lengths and noise levels, 4) the impact of receiver's detection efficiency, and 5) an evaluation of the KPS's performance.

\vspace{-8pt}
\subsection{Effect of Data Transmission Speed}
Data transmission speed is a critical statistic in a QKD-based microgrid. A speed larger than the key generation speed can result in the exhaustion of key bits in a key pool, eventually causing the failure of data communication. 

We used Wireshark, a free and open-source packet analyzer, to monitor traffic in the system. Specifically, two types of packets were captured: the packets sent from the RTDS (GTNETx2 \#2) to the MGCC and from the MGCC to the RTDS (GTNETx2 \#1). The transmission speed of the two types of packets were set as the same. Namely, once there was a packet received by the MGCC, a packet was sent out from the MGCC.

\begin{figure}[t]
\centering
\includegraphics[width=0.48\textwidth]{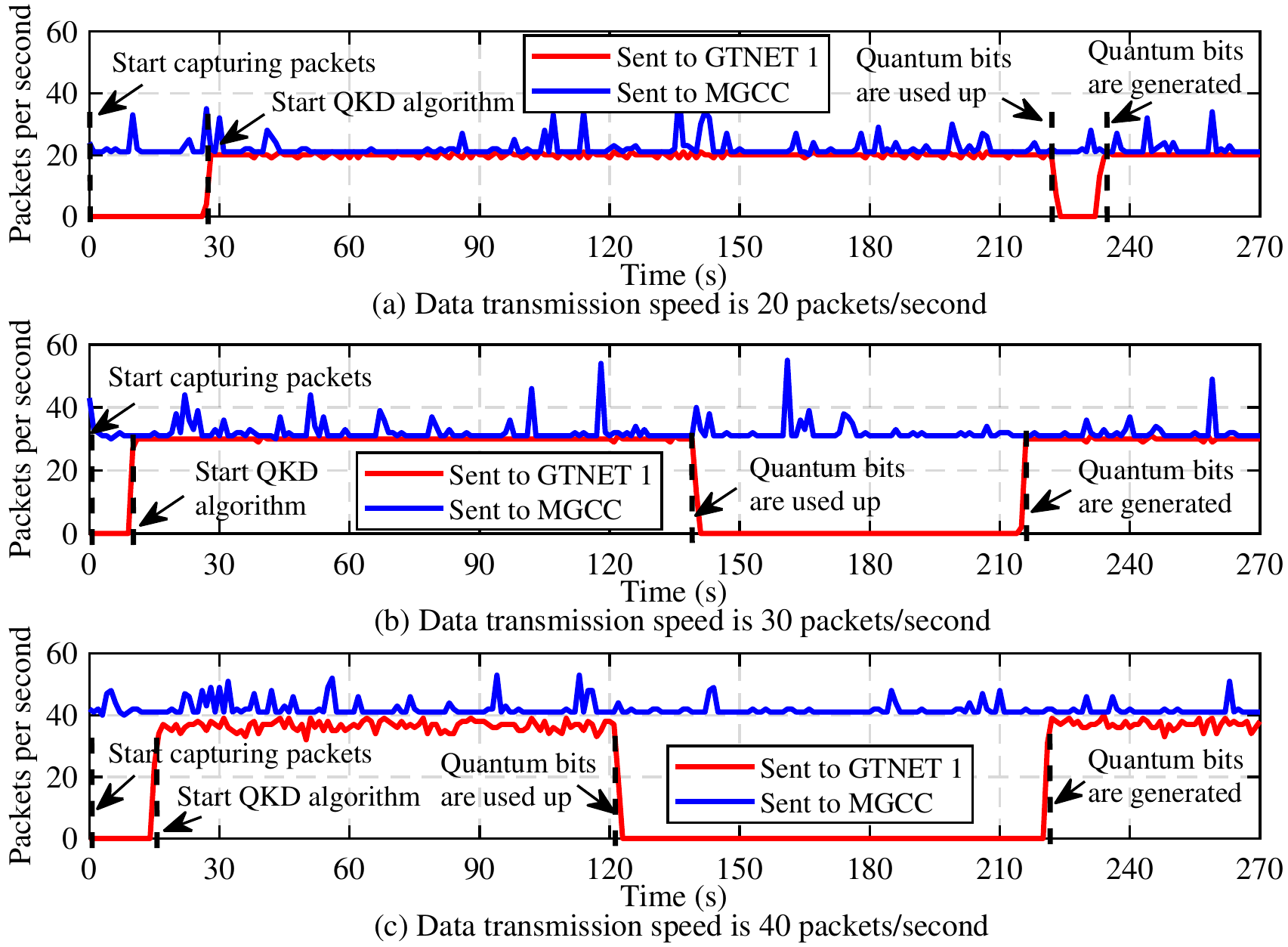}
\vspace{-6pt}
\caption{\label{fig:packetcapture}Traffic monitoring under different data transmission speeds.}
\vspace{-12pt}
\end{figure}

The impact of the data transmission speed is illustrated in Fig.~\ref{fig:packetcapture}, where the fiber length $L$ (between the MGCC and the local controller) is set at 50 km. The other parameters are the same as those in Table~\ref{tab:paraQKD}. Each packet sent from the MGCC to the RTDS consists of 64 binary bits, meaning that 64 key bits are consumed from the key pool when a packet is sent out.

From Fig.~\ref{fig:packetcapture}, it can be observed that:
\begin{itemize}
    \item The data transmission speed has a large impact on the QKD-based microgrid. With the setting in Fig.~\ref{fig:packetcapture}, a speed larger than 20 packets/second will lead to the exhaustion of key bits in the key pool.
    \item The larger the data transmission speed, the sooner the quantum bits will be consumed. With the setting in Fig.~\ref{fig:packetcapture}, for a speed of 40 packets/second, the exhaustion lasts around 100 seconds within the key generation period. This long shortage can cause serious damage to microgrid operations, as there is no key in the key pool for the encryption and authentication of data messages.
\end{itemize}

\vspace{-8pt}
\subsection{Impact of Cyberattacks on the Microgrid}
For either a classical communication or a quantum communication system during the exhaustion of key bits, the security of the system can be easily broken by an adversary using quantum computers, leading to insecurity in both the encryption and authentication of data messages. The data messages sent from the MGCC to local controllers can thus be intercepted, decrypted, falsified, re-encrypted, and re-sent to local controllers by an adversary without being detected.

\begin{figure}[t]
\centering \subfigure[Voltage response of bus 1 before and after the attack without QKD]
{\includegraphics[width=0.48\textwidth,angle=0]{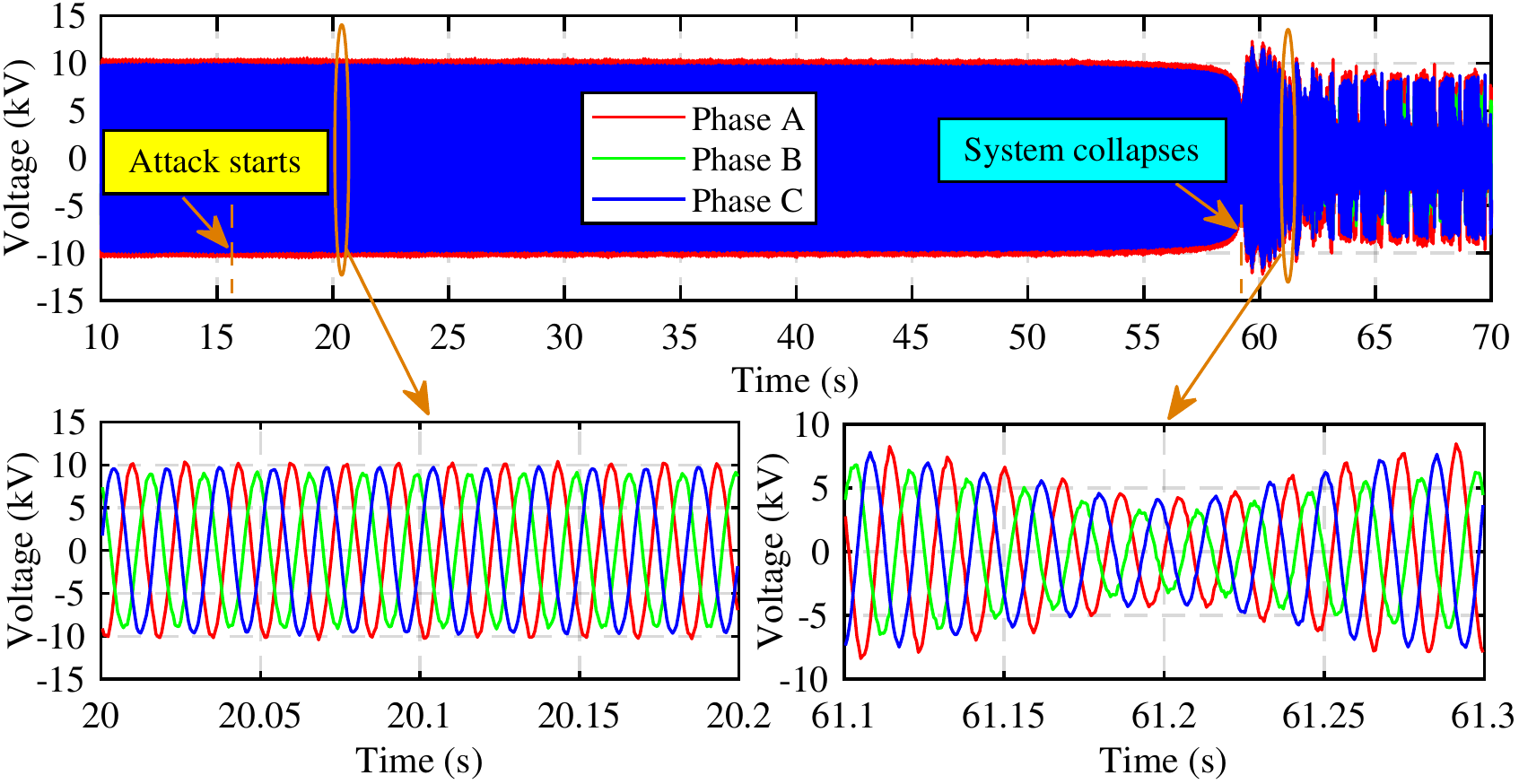}}
\centering \subfigure[Voltage response of bus 1 with QKD]
{\includegraphics[width=0.48\textwidth,angle=0]{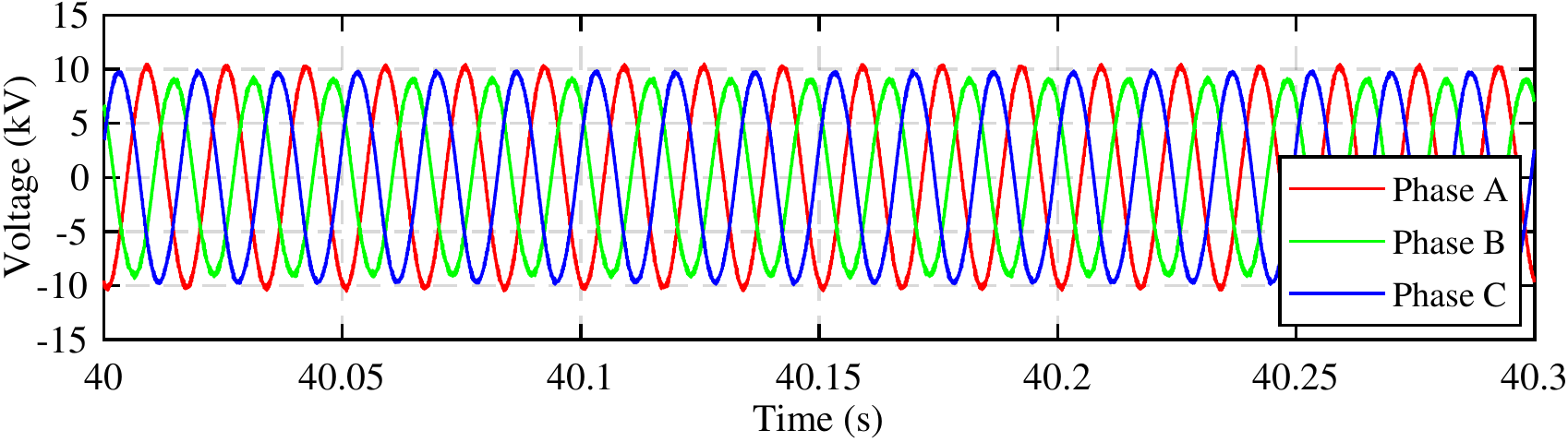}}
\vspace{-6pt}
\caption{Voltage response of bus 1 with and without QKD.}
\label{fig:Bus1Voltage}
\vspace{-12pt}
\end{figure}

To test the impact of a malicious control signal on the microgrid system, the real power reference of the P-Q control for the battery is changed from the initial value, 0, to -6 MW at time $t=16$ s during the islanded mode. The voltage response of bus 1 before and after the attack is illustrated in Fig.~\ref{fig:Bus1Voltage} (a). It shows that, 1) the magnitude of voltage gradually decreases; 2) the frequency also decreases; and 3) at time $t=59$ s, the system eventually collapses. However, if QKD is employed and there are enough key bits in the key pool, it will be impossible to break the encryption or authentication due to the unconditional security of QKD, and thus no malicious data can be injected. The normal voltage response of bus 1 in a QKD-based microgrid is illustrated in Fig.~\ref{fig:Bus1Voltage} (b).

\vspace{-8pt}
\subsection{Evaluation of Key Generation Speed under Different Fiber Lengths and Noise Levels}
The speed of quantum key generation determines the maximum data transmission speed in a QKD-based microgrid. The larger the key generation speed, the higher the maximum data transmission speed. However, it was unclear which levels of key generation speed the QKD system could provide for the microgrid under different conditions. In this subsection, an evaluation of key generation speed under different fiber lengths $L$s and noise levels $e_{mis}$s, is provided. The noise can be either natural or caused by an adversary. A strong attack on the quantum optic equipment leads to a large $e_{mis}$.

\begin{figure}[t]
\centering
\includegraphics[width=0.38\textwidth]{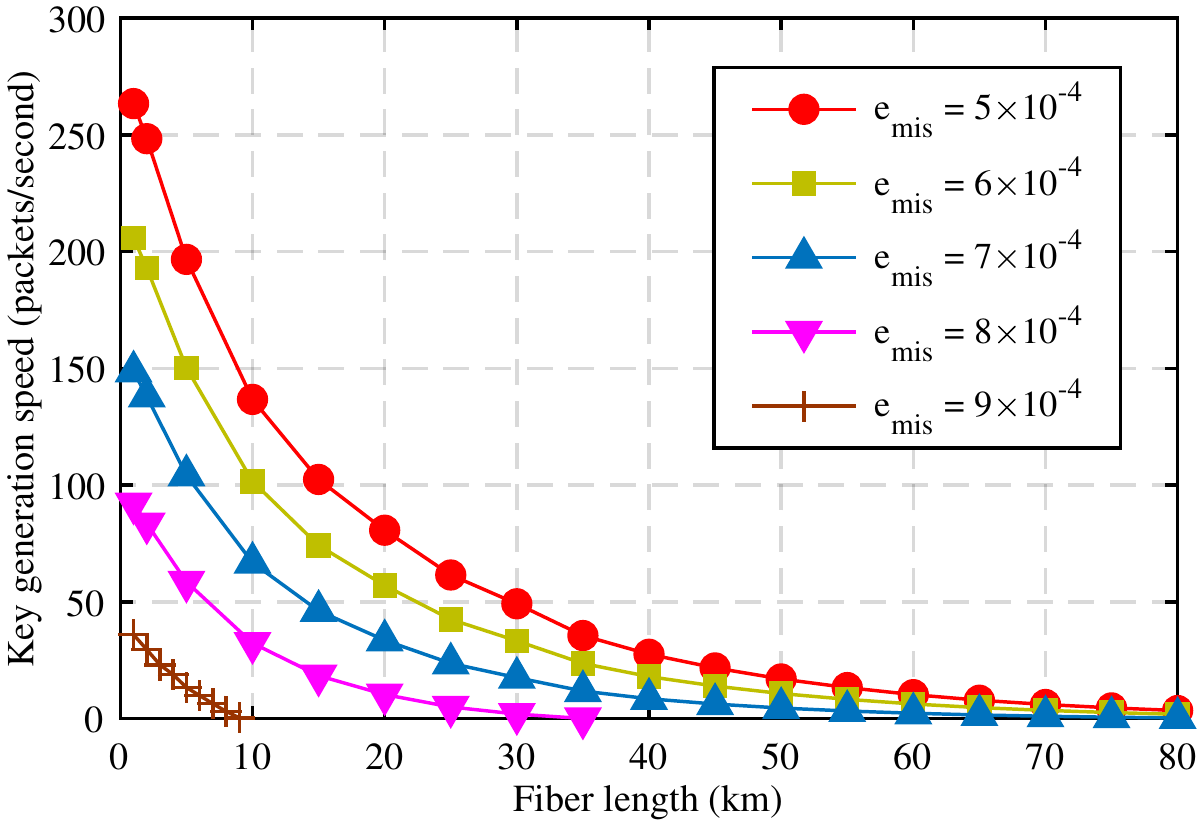}
\vspace{-8pt}
\caption{\label{fig:AttDisSpeed}Quantum key generation speeds under different $L$s and $e_{mis}$s.}
\vspace{-8pt}
\end{figure}

The real-time simulation results are given in Fig.~\ref{fig:AttDisSpeed}, where $L$ is set from 1 km to 80 km, $e_{mis}$ is set from 5$\times$10$^{-4}$ to 9$\times$10$^{-4}$ with a step of 1$\times$10$^{-4}$, and each packet consists of 64 binary bits. The other parameters are the same as those in Table~\ref{tab:paraQKD}. Key generation speed is calculated as the fraction of the generated key's size $\ell$ (see (\ref{equ:length})) and the time required.

It can be observed that:

\begin{itemize}
    \item A small $L$ exhibits great superiority over a large $L$ under the same $e_{mis}$, which gives valuable insights that the MGCC and the local controller should be close to each other in a QKD-based microgrid.
    \item The key generation speed is sufficient with a small $L$ and a small $e_{mis}$. But, it decreases dramatically when $e_{mis}$ increases. A proper strategy therefore has to be carried out to improve the system's cyberattack resilience.
    \item Importantly, Fig.~\ref{fig:AttDisSpeed} gives valuable resources on which levels the data transmission speed should be set at under different $L$s and $e_{mis}$s. With the setting in Fig.~\ref{fig:AttDisSpeed}, any data transmission speed that is below the corresponding curve (with regards to a certain $e_{mis}$) in Fig.~\ref{fig:AttDisSpeed}, will have sufficient key bits in the key pool under that $e_{mis}$.
\end{itemize}

\vspace{-10pt}
\subsection{The Impact of Receiver's Detection Efficiency}
The detection efficiency of the receiver, $\eta_{Bob}$, is critical in a QKD system. Detection efficiency refers to the probability that the receiver can successfully detect the photons, which is largely determined by the quality of the detection devices.

\begin{figure}[t]
\centering
\includegraphics[width=0.38\textwidth]{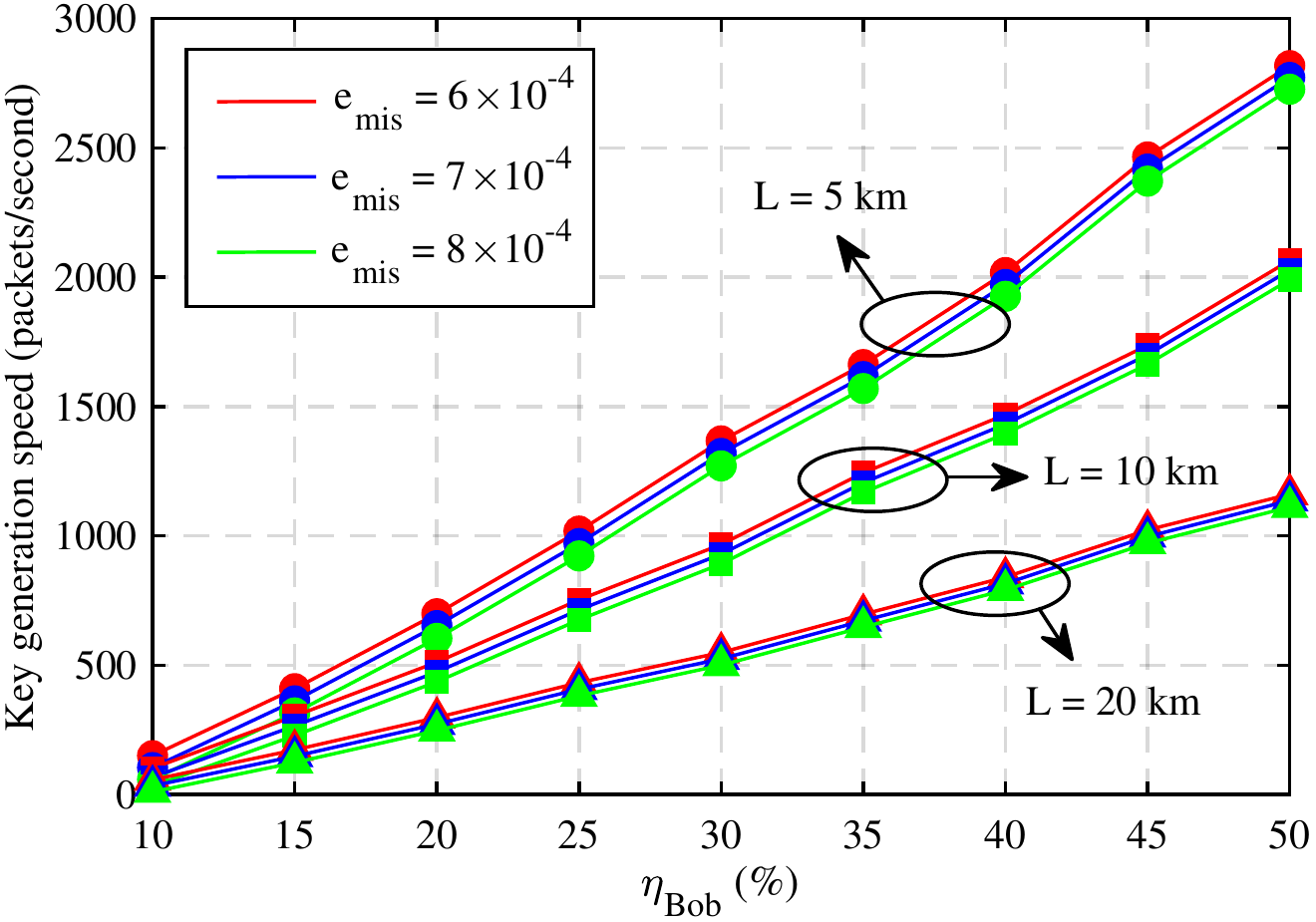}
\vspace{-8pt}
\caption{\label{fig:EfficiencyBob}Quantum key generation speeds under different $\eta_{Bob}$s.}
\vspace{-12pt}
\end{figure}

The impact of $\eta_{Bob}$ is evaluated in our real-time testbed. The results are illustrated in Fig.~\ref{fig:EfficiencyBob}, where $L$ is set at 5 km, 10 km, and 20 km, respectively; $e_{mis}$ is set at 6$\times$10$^{-4}$, 7$\times$10$^{-4}$, and 8$\times$10$^{-4}$, respectively; and $\eta_{Bob}$ is from 10\% to 50\% with a step of 5\%. The other parameters are the same as in Table~\ref{tab:paraQKD}.

It can be seen that $\eta_{Bob}$ has a significant impact on key generation speed. With a given $L$ and a given $e_{mis}$, a small increase of $\eta_{Bob}$ results in a great improvement of the speed. This indicates that it is worth improving the quality of detection devices in a QKD-based microgrid.

\vspace{-10pt}
\subsection{Evaluation of KPS Performance}
The performance of the presented KPS strategy is evaluated in our testbed. In this test case, two key pools are established in the quantum algorithm, and each stores its quantum key bits separately. The QKD parameters for the two key pools are set as the same except that $e_{mis}$ for key pool \#1 is 8$\times$10$^{-4}$ to simulate a strong attack, while $e_{mis}$ for key pool \#2 is 5$\times$10$^{-4}$ for a weak attack. The data transmission speed is set at 100 packets/second, where each packet consists of 64 bits.

For the KPS strategy, the threshold is set at 5,000 bits for key pool \#1, meaning that once the number of key bits in key pool \#1 is lower than 5,000, a given number (which is set at 20,000) of key bits will be shared from key pool \#2.

The comparison results of the numbers of key bits in key pools \#1 and \#2 with and without KPS are illustrated  in Fig.~\ref{fig:keypoolsize_KPS}. It can be observed that:

\begin{figure}[t]
\centering
\includegraphics[width=0.48\textwidth]{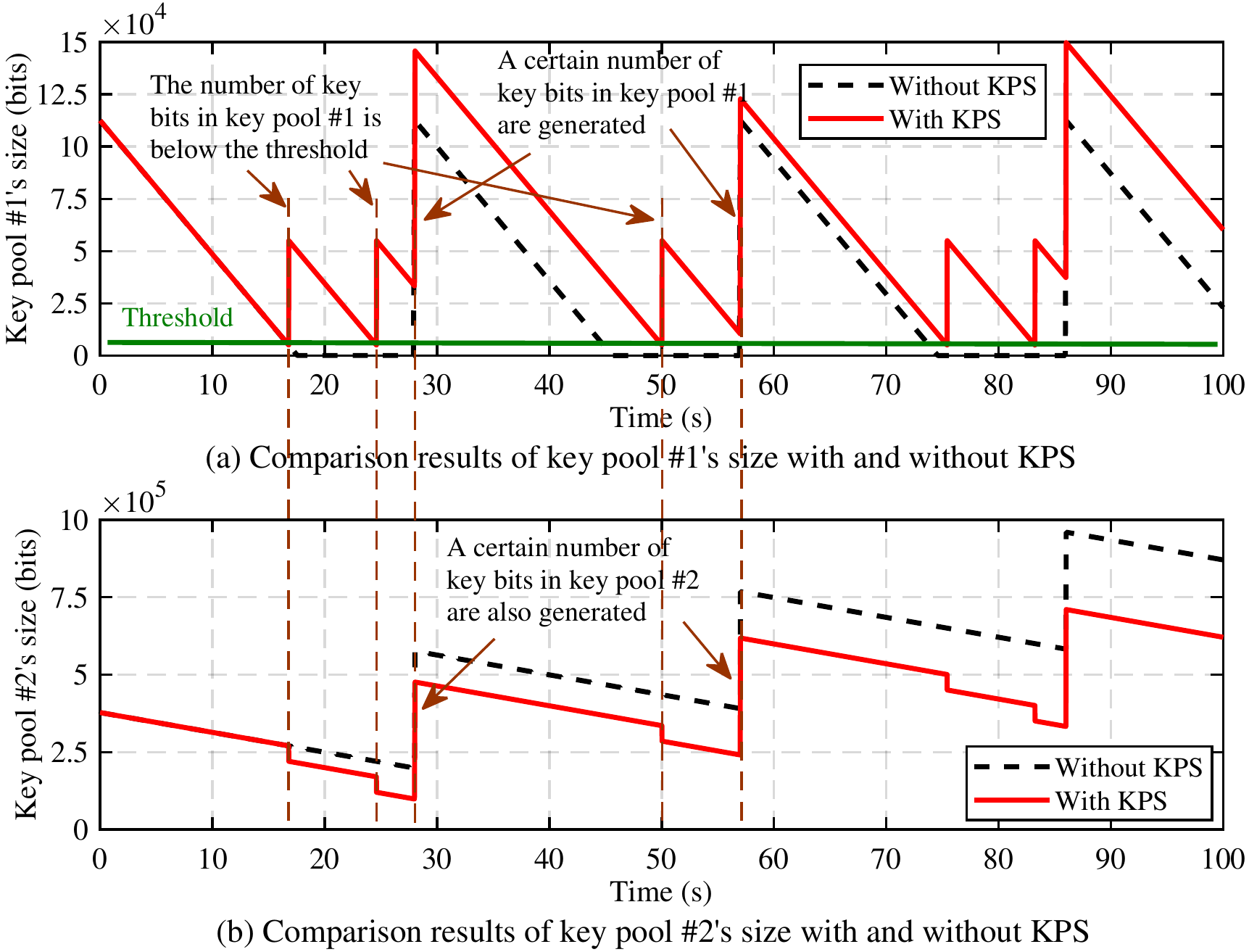}
\vspace{-8pt}
\caption{\label{fig:keypoolsize_KPS}Comparison results of the numbers of key bits in key pools \#1 and \#2 with and without KPS.}
\vspace{-12pt}
\end{figure}

\begin{itemize}
    \item Without KPS, there is a shortage of key bits in key pool \#1. For instance, at time $t=17.56$ s, the key bits in key pool \#1 are used up (see the black dashed line in Fig.~\ref{fig:keypoolsize_KPS} (a)), and the shortage lasts around 10.5 s until a certain number of key bits are generated. Meanwhile, the key bits in key pool \#2 do not have shortage issues (see the black dashed line in Fig.~\ref{fig:keypoolsize_KPS} (b)).
    \item With KPS, the shortage issues of key pool \#1 are well addressed. At time $t=16.79$ s, the number of key bits in key pool \#1 is below the threshold, and immediately 20,000 key bits are added (see the red solid line in Fig.~\ref{fig:keypoolsize_KPS} (a)). Meanwhile, 20,000 key bits are deducted from key pool \#2 (see the red solid line in Fig.~\ref{fig:keypoolsize_KPS} (b)). But this does not affect the normal operation of key pool \#2, as the minimum number of key bits in key pool \#2 is still above the threshold.
\end{itemize}

\vspace{-8pt}
\section{Conclusion}

This paper presents a real-time QKD-enabled microgrid testbed implemented in RTDS. This testbed provides a realistic cyber-physical testing environment in real time with a simulated QKD algorithm integrated. This is an important step towards constructing a real QKD system in microgrid in practice. With this testbed, more research work could be done in the future. Some examples include exploiting the feasibility of more advanced and practical QKD protocols for microgrids, evaluating the QKD-enabled microgrid's performance under more scenarios, and developing methods to further enhance the cyberattack resilience of the QKD-enabled microgrid.

\ifCLASSOPTIONcaptionsoff
  \newpage
\fi
 
\bibliographystyle{IEEEtran}
\bibliography{mybibfile}

\begin{thebibliography}{10}
\providecommand{\url}[1]{#1}
\csname url@samestyle\endcsname
\providecommand{\newblock}{\relax}
\providecommand{\bibinfo}[2]{#2}
\providecommand{\BIBentrySTDinterwordspacing}{\spaceskip=0pt\relax}
\providecommand{\BIBentryALTinterwordstretchfactor}{4}
\providecommand{\BIBentryALTinterwordspacing}{\spaceskip=\fontdimen2\font plus
\BIBentryALTinterwordstretchfactor\fontdimen3\font minus
  \fontdimen4\font\relax}
\providecommand{\BIBforeignlanguage}[2]{{%
\expandafter\ifx\csname l@#1\endcsname\relax
\typeout{** WARNING: IEEEtran.bst: No hyphenation pattern has been}%
\typeout{** loaded for the language `#1'. Using the pattern for}%
\typeout{** the default language instead.}%
\else
\language=\csname l@#1\endcsname
\fi
#2}}
\providecommand{\BIBdecl}{\relax}
\BIBdecl

\bibitem{farrokhabadi2019microgrid}
M.~Farrokhabadi, C.~A. Canizares, J.~W. Simpson-Porco, E.~Nasr, L.~Fan,
  P.~Mendoza-Araya, R.~Tonkoski, U.~Tamrakar, N.~D. Hatziargyriou, D.~Lagos
  \emph{et~al.}, ``Microgrid stability definitions, analysis, and examples,''
  \emph{IEEE Transactions on Power Systems}, 2019.

\bibitem{ren2016enabling}
L.~Ren, Y.~Qin, B.~Wang, P.~Zhang, P.~B. Luh, and R.~Jin, ``Enabling resilient
  microgrid through programmable network,'' \emph{IEEE Transactions on Smart
  Grid}, vol.~8, no.~6, pp. 2826--2836, 2016.

\bibitem{van2014quantum}
R.~Van~Meter, \emph{Quantum networking}.\hskip 1em plus 0.5em minus 0.4em\relax
  John Wiley \& Sons, 2014.

\bibitem{banik2019compact}
S.~Banik, A.~Bogdanov, and F.~Regazzoni, ``Compact circuits for combined {AES}
  encryption/decryption,'' \emph{Journal of Cryptographic Engineering}, vol.~9,
  no.~1, pp. 69--83, 2019.

\bibitem{liskov2019enrich}
M.~D. Liskov, J.~D. Guttman, J.~D. Ramsdell, P.~D. Rowe, and F.~J. Thayer,
  ``Enrich-by-need protocol analysis for {Diffie-H}ellman,'' in
  \emph{Foundations of Security, Protocols, and Equational Reasoning}.\hskip
  1em plus 0.5em minus 0.4em\relax Springer, 2019, pp. 135--155.

\bibitem{coutinho1999mathematics}
S.~C. Coutinho, \emph{The mathematics of ciphers: number theory and {RSA}
  cryptography}.\hskip 1em plus 0.5em minus 0.4em\relax AK Peters/CRC Press,
  1999.

\bibitem{mccurley1990discrete}
K.~S. McCurley, ``The discrete logarithm problem,'' in \emph{AMS Proc. Symp.
  Appl. Math}, vol.~42, 1990, pp. 49--74.

\bibitem{shor1994algorithms}
P.~W. Shor, ``Algorithms for quantum computation: {D}iscrete logarithms and
  factoring,'' in \emph{Proceedings 35th annual symposium on foundations of
  computer science}.\hskip 1em plus 0.5em minus 0.4em\relax Ieee, 1994, pp.
  124--134.

\bibitem{yan2019logarithm}
S.~Y. Yan, ``Logarithm based cryptography,'' in \emph{Cybercryptography:
  Applicable Cryptography for Cyberspace Security}.\hskip 1em plus 0.5em minus
  0.4em\relax Springer, 2019, pp. 287--341.

\bibitem{lara2019trends}
P.~D.~M. Lara, D.~A. Maldonado-Ruiz, S.~D.~A. D{\'\i}az \emph{et~al.}, ``Trends
  on computer security: {C}ryptography, user authentication, denial of service
  and intrusion detection,'' \emph{arXiv preprint arXiv:1903.08052}, 2019.

\bibitem{fano2019quantum}
G.~Fano and S.~Blinder, ``Quantum chemistry on a quantum computer,'' in
  \emph{Mathematical Physics in Theoretical Chemistry}.\hskip 1em plus 0.5em
  minus 0.4em\relax Elsevier, 2019, pp. 377--400.

\bibitem{wright2019benchmarking}
K.~Wright, K.~Beck, S.~Debnath, J.~Amini, Y.~Nam, N.~Grzesiak, J.-S. Chen,
  N.~Pisenti, M.~Chmielewski, C.~Collins \emph{et~al.}, ``Benchmarking an
  11-qubit quantum computer,'' \emph{arXiv preprint arXiv:1903.08181}, 2019.

\bibitem{kovachy2015quantum}
T.~Kovachy, P.~Asenbaum, C.~Overstreet, C.~Donnelly, S.~Dickerson,
  A.~Sugarbaker, J.~Hogan, and M.~Kasevich, ``Quantum superposition at the
  half-metre scale,'' \emph{Nature}, vol. 528, no. 7583, p. 530, 2015.

\bibitem{bengtsson2017geometry}
I.~Bengtsson and K.~{\.Z}yczkowski, \emph{Geometry of quantum states: {A}n
  introduction to quantum entanglement}.\hskip 1em plus 0.5em minus 0.4em\relax
  Cambridge university press, 2017.

\bibitem{orus2019quantum}
R.~Or{\'u}s, S.~Mugel, and E.~Lizaso, ``Quantum computing for finance:
  {O}verview and prospects,'' \emph{Reviews in Physics}, p. 100028, 2019.

\bibitem{hong2020quantum}
C.~Hong, J.~Jang, J.~Heo, and H.-J. Yang, ``Quantum digital signature in a
  network,'' \emph{Quantum Information Processing}, vol.~19, no.~1, p.~18,
  2020.

\bibitem{bani2019online}
A.~Bani-Hani, M.~Majdalweieh, and A.~AlShamsi, ``Online authentication methods
  used in banks and attacks against these methods,'' \emph{Procedia Computer
  Science}, vol. 151, pp. 1052--1059, 2019.

\bibitem{cobourne2011quantum}
S.~Cobourne \emph{et~al.}, ``Quantum key distribution protocols and
  applications,'' \emph{Surrey TW20 0EX, England}, 2011.

\bibitem{trauzettel2007spin}
B.~Trauzettel, D.~V. Bulaev, D.~Loss, and G.~Burkard, ``Spin qubits in graphene
  quantum dots,'' \emph{Nature Physics}, vol.~3, no.~3, p. 192, 2007.

\bibitem{pirandola2019}
S.~Pirandola, U.~Andersen, L.~Banchi, M.~Berta \emph{et~al.}, ``Advances in
  quantum cryptography,'' \emph{arXiv preprint arXiv:1906.01645}, 2019.

\bibitem{lim2014concise}
C.~C.~W. Lim, M.~Curty, N.~Walenta, F.~Xu, and H.~Zbinden, ``Concise security
  bounds for practical decoy-state quantum key distribution,'' \emph{Physical
  Review A}, vol.~89, no.~2, p. 022307, 2014.

\bibitem{nagata2019no}
K.~Nagata, T.~Nakamura \emph{et~al.}, ``No-cloning theorem, {Kochen-S}pecker
  theorem, and quantum measurement theories,'' \emph{International Journal of
  Theoretical Physics}, vol.~58, no.~6, pp. 1845--1853, 2019.

\bibitem{fung2010practical}
C.-H.~F. Fung, X.~Ma, and H.~Chau, ``Practical issues in
  quantum-key-distribution postprocessing,'' \emph{Physical Review A}, vol.~81,
  no.~1, p. 012318, 2010.

\bibitem{eraerds2010quantum}
P.~Eraerds, N.~Walenta, M.~Legr{\'e}, N.~Gisin, and H.~Zbinden, ``Quantum key
  distribution and 1 {Gbps} data encryption over a single fibre,'' \emph{New
  Journal of Physics}, vol.~12, no.~6, p. 063027, 2010.

\bibitem{chen2016report}
L.~Chen, S.~Jordan, Y.-K. Liu, D.~Moody, R.~Peralta, R.~Perlner, and
  D.~Smith-Tone, \emph{Report on post-quantum cryptography}.\hskip 1em plus
  0.5em minus 0.4em\relax US Department of Commerce, National Institute of
  Standards and Technology, 2016.

\bibitem{aes}
``{AES-NI SSL Performance: A study of AES-NI acceleration using LibreSSL,
  OpenSSL},'' [Online available]:
  https://calomel.org/aesni{\_ssl\_performance.html}.

\bibitem{onyinyechi2015real}
N.~Onyinyechi, ``Real time simulation of a microgrid system with distributed
  energy resources,'' 2015.

\end{thebibliography}

\end{document}